\documentclass[runningheads]{llncs}
\usepackage{graphicx}
\usepackage{amsmath}
\usepackage{float}
\usepackage{caption}
\usepackage{dsfont}
\usepackage{subcaption}
\usepackage{hyperref}
\begin{document}
\title{Contrastive-Adversarial and Diffusion: Exploring pre-training and fine-tuning strategies for sulcal identification}
\titlerunning{ }
%
\author{Michail Mamalakis\inst{1,2} \and
Héloïse de Vareilles \inst{1} \and
Shun-Chin Jim Wu \inst{1} \and
Ingrid Agartz \inst{6} \and
Lynn Egeland Mørch-Johnsen \inst{4,5} \and
Jane  Garrison\inst{4} \and
Jon Simons \inst{4} \and
Pietro Lio\inst{2} \and
John Suckling\inst{1} \and
Graham Murray\inst{1}}
\authorrunning{M. Mamalakis et al.}

\institute{Department of Psychiatry, University of Cambridge, Cambridge, UK. \and Department of Computer Science and Technology, Computer Laboratory, University of Cambridge, Cambridge, UK. \and Department of Psychology, University of Cambridge, Cambridge, UK. \and Norment, Division of Mental Health and Addiction, Oslo University Hospital, Institute of Clinical Medicine, University of Oslo, Oslo, Norway. \and Department of Psychiatry and Department of Clinical Research, Østfold Hospital, Grålum, Norway. \and Department of Psychiatric Research, Diakonhjemmet Hospital, Oslo, Norway}
\maketitle

\begin{abstract}
In the last decade, computer vision has witnessed the establishment of various training and learning approaches. Techniques like adversarial learning, contrastive learning, diffusion denoising learning, and ordinary reconstruction learning have become standard, representing state-of-the-art methods extensively employed for fully training or pre-training networks across various vision tasks. The exploration of fine-tuning approaches has emerged as a current focal point, addressing the need for efficient model tuning with reduced GPU memory usage and time costs while enhancing overall performance, as exemplified by methodologies like low-rank adaptation (LoRA). Key questions arise: which pre-training technique yields optimal results—adversarial, contrastive, reconstruction, or diffusion denoising? How does the performance of these approaches vary as the complexity of fine-tuning is adjusted?
This study aims to elucidate the advantages of pre-training techniques and fine-tuning strategies to enhance the learning process of neural networks in independent identical distribution (IID) cohorts. We underscore the significance of fine-tuning by examining various cases, including full tuning, decoder tuning, top-level tuning, and fine-tuning of linear parameters using LoRA. Systematic summaries of model performance and efficiency are presented, leveraging metrics such as accuracy, time cost, and memory efficiency.
To empirically demonstrate our findings, we focus on a multi-task segmentation/classification challenge involving the paracingulate sulcus (PCS) using different 3D Convolutional Neural Network (CNN) architectures by using the TOP-OSLO cohort comprising 596 subjects.
\end{abstract}

\section{Introduction}
Segmenting small regions, such as the paracingulate sulcus (PCS), presents a distinct challenge due to its secondary subregion status, comprising a small fraction of total sulcal regions. The inherent variability in shape, size, and location, coupled with uncertainty in PCS anatomy, poses difficulties for well-trained models to generalize effectively across the same (IID) or diverse cohorts (out-of-distribution; OOD). 
\par Multi-task learning, commonly employed to tackle complex vision tasks or address imbalanced classes \cite{2}, provides a framework for simultaneously training and exploring interactions between different hypotheses within the context of a generalized task. Existing research suggests the potential for improved Convolutional Neural Network (CNN) performance in multi-task scenarios compared to single, well-defined computer vision tasks \cite{3,5}. However, the challenge lies in identifying specific tasks that yield optimal results for a given generalized objective, with common combinations involving segmentation and classification tasks \cite{3,5}. Combining pre-training and fine-tuning tasks has proven effective for enhancing model generalization in intricate segmentation tasks \cite{1}. 
\par Pre-training methods, such as diffusion denoising, adversarial, contrastive, or reconstruction learning, play a vital role. Diffusion models excel in generating synthetic images to balance highly unbalanced cohorts in various medical domains, brain imaging and multi-modal computer vision tasks \cite{5,11,13}. 
\par Adversarial learning, known for synthesizing images by learning specific distributions, has seen innovations, including hybrid approaches integrating distribution conditions from diffusion models \cite{16}. Multi-context, explored in adversarial learning within medical imaging \cite{18}, demonstrates enhanced generalization and accuracy in pre-training and main-training for segmentation \cite{1}. 
\par Contrastive learning, a self-supervised technique, explores cohort characteristics through negative-positive probability pairs \cite{22}. Hybrid ideas, combining negative-positive sample sets with diffusion denoising techniques, have been proposed \cite{19}. Strategies like auxiliary guidance and the contrastive diffusion approach enhance input-output correspondence, improving clinical reliability. Contrastive learning serves various purposes, acting as a supervised method for anatomical interpretability, weakly-supervised for disease detection, and semi-supervised for generalizing specific computer vision tasks \cite{20,23,15}. \par Reconstruction techniques diversely increase input resolution. Approaches combining multi-modality and denoising in both k-space and image-space show promise in in-vivo experiments \cite{25}. In some instances, reconstruction serves as a self-supervised method, learning the distribution in an encoder-decoder architecture, and subsequently segmenting the region of interest through fine-tuning techniques \cite{26}.
\par On another front, fine-tuning strategies and domain adaptation techniques aim to adjust pre-trained networks for specific tasks and adapt them to out-of-distribution (OOD) cohorts \cite{30}. Moreover, fine-tuning, particularly in an identical distribution (IID) cohort, is recommended to prevent overfitting, ensuring high performance, and improving generalization to potential OOD cohorts \cite{28,31}. Fine-tuning strategies vary based on which parameters are 'frozen' (not trained) and 'hot' (retrained). Low-rank adaptation (LoRA), a well-established fine-tuning method in the large language model domain \cite{34}, has proven effective in this context.
\par Several pivotal questions arise based on the pre-training fine-tuning strategy: Which approach yields optimal results—adversarial, contrastive, reconstruction, or diffusion denoising? How does the performance of these methods evolve as we adjust the complexity of fine-tuning? Can a combined self-supervised pre-training and fine-tuning strategy effectively address the challenges posed by high uncertainty and complexity in computer vision tasks, particularly in the segmentation of small regions with uncertain locations? 
\par This research explores a multi-task challenge involving the segmentation and classification of the paracingulate sulcus (PCS). It employs diverse 3D transformer convolutional neural network architectures and leverages a thoroughly annotated cohort of 596 subjects from the TOP-OSLO study \cite{morch-johnsen}.The primary objectives of this preliminary study are to elucidate the advantages of diverse pre-training techniques—adversarial learning, contrastive learning, diffusion denoising learning, and reconstruction learning—coupled with fine-tuning strategies to enhance the learning process of neural networks in IID cohorts with the potential to benefit the OOD performance. We underscore the importance of fine-tuning learning through an exploration of different scenarios, encompassing full retraining, decoder training, top-level training, and the fine-tuning of linear parameters using Low-Rank Adaptation (LoRA). A systematic summarization of model performance and efficiency is conducted, with metrics such as accuracy, time cost, and memory efficiency serving as key evaluative criteria. Through this comprehensive analysis, we aim to shed light on the most effective approaches for addressing the intricate challenges associated with the segmentation of small regions, particularly the PCS, in medical image analysis.
\section{Methods}
\subsection{Overview and experiment}
\begin{figure}
\caption{The proposed strategy for complex multi-task learning computer vision tasks.}
  \label{x4}
      \medskip
\centerline{
\includegraphics[scale=.35]{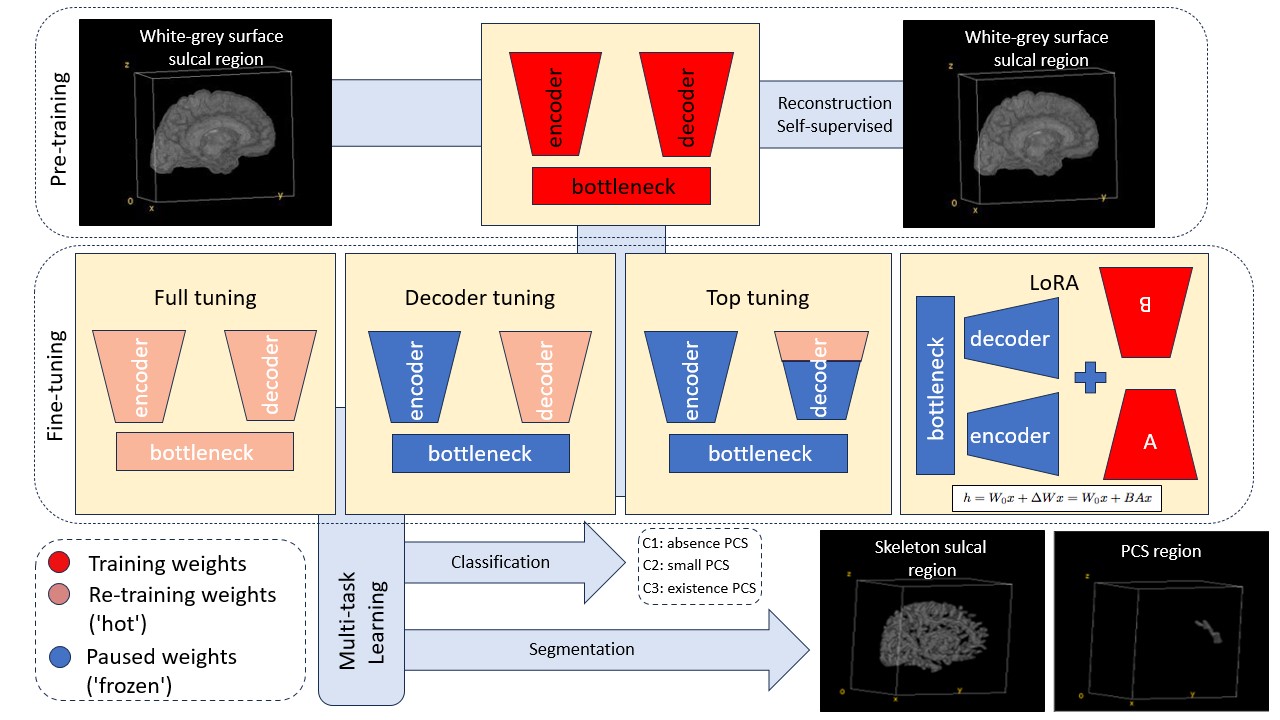}
}
\end{figure}

Our methodology integrates pre-training and fine-tuning simulations to tackle the complex multi-task computer vision challenge involving paracingulate sulcus (PCS) segmentation and patient classification into three categories: absence of PCS, existence of PCS, and a small PCS region (Fig. \ref{x4}). For the pre-training's self-supervised reconstruction task, we use the white-grey surface matter of the subjects' brains as an input, with a specific focus on the left hemisphere, where classification results exhibit superior performance (\cite{our}). Focused on fine-tuning, this study emphasizes a multi-task segmentation and classification challenge centered around the paracingulate sulcus (PCS), employing diverse 3D transformer convolutional neural network architectures. Leveraging a comprehensive cohort of 596 subjects from the TOP-OSLO study \cite{morch-johnsen}, we partition the dataset into training (70\%), validation (20\%), and testing (10\%) sets. In the fine-tuning simulation for the segmentation task, we address a two-label segmentation task involving the skeleton sulcal and PCS regions, utilizing the white-grey surface matter of the patient (Fig. \ref{x4}) extracted from T1-weighted MRI. Our image preprocessing aligns with the outlined steps in \cite{our}. Both pre-training and fine-tuning are conducted on the training and validation cohorts, and the networks are evaluated on the unseen testing cohort. We used an NVIDIA A100-SXM-80GB GPU, and the code will be publicly accessible through github.
\subsection{Pre-training methods}
In the pre-training phase, the networks undergo self-supervised learning, aiming to reconstruct the input as an output. This approach ensures that the encoder acquires information about the reduced dimensionality space in the latent space, which is then decoded back to the vision space, reconstructing the input. Our pre-training strategies include adversarial \cite{18}, contrastive \cite{rec}, simple reconstruction \cite{rec}, and diffusion denoising learning \cite{diff}.
For the simple reconstruction, the cost function employed is as follows:
\begin{equation}
SSIM(x,y)= 1- \frac{(2 \mu_x \mu_y + c_1)(2 \sigma_{xy}+c_2)}{(\mu_x^2+\mu_y^2+c_1)(\sigma_x^2+\sigma_y^2+c_2)}
\label{1}
\end{equation}
\begin{equation}
loss_{rec}(x,y)=SSIM(x,y)
\label{2}
\end{equation}
where $x$ the prediction, $y$ the ground truth, $\mu_(x)$ the average of $x$, $\sigma_x$ the variance of x, $\sigma_{xy}$ the covariance of $x$ and $y$, and $c_1$ and $c_2$ two variables to stabilize the division with weak denominator (\cite{ssim}).
The total loss function was given by:
\begin{equation}
loss_{total}(x,y)=loss_{rec}(x,y) 
\label{3}
\end{equation}
For the adversial learning the cost function was:
\begin{equation}
loss_{disc}(x_{fake},y_{fake},x_{real},y_{real})= \frac{CE(x_{fake},y_{fake})+CE(x_{real},y_{real})}{2}
\label{4}
\end{equation}
\begin{equation}
loss_{gen}(x_{fake},y_{real})= CE(x_{fake},y_{real})
\label{5}
\end{equation}
\begin{equation}
loss_{rec}(x,y)=SSIM(x,y)
\label{6}
\end{equation}
where $x_{fake}$ is the output of the discriminator when is feeding from the prediction of generator, $x_{real}$ is the prediction of the discriminator. The $y_{real}$ is the ground truth anotated as real image with 1 and $y_{fake}$ the ground truth annotated as fake image with 0. Again $x$ and $y$ the reconstruction predictions and the groundtruth. 
The discriminator loss is backpropagated every $n+1$ epochs, while the generator and the total loss, $loss_{total}$, are updated every $n$ epochs.
Moving to contrastive learning:
\begin{equation}
loss_{i,j}^{con}(x)= -log \frac{exp(sim(x_i,x_j)/ \tau)}{\sum_{k=1}^{2N}\mathds{1}_{[k\neq i]} exp(sim(x_i,x_k)/ \tau)}
\label{7}
\end{equation}
\begin{equation}
loss_{rec}(x,y)=SSIM(x,y)+loss_{i,j}^{con}(x)
\label{8}
\end{equation}
where $\mathds{1}_{[k\neq i]}$ is an indicator function evaluating to 1 if $k\neq i$ and $\tau$ denotes a temperature parameter. The final loss is computed across all positive pairs, both $(i,j)$ and $(j,i)$, in a mini-batch (\cite{con2}). 

For diffusion denoising the cost function compute based on eq. \ref{2}.
Regarding the simple reconstruction, contrastive, and the generator network of the adversarial learning, we employed the Swin-UNETR \cite{rec}, with a 24 feature size. For the diffusion denoising, we opted a Diffusion Denoising U-NET network (DD-UNET, \url{https://github.com/Project-MONAI/GenerativeModels}). This model consisted of 3 levels with the number of channels set to 32, 32, and 64. Attention was applied only in the third level, with a number of attention heads set to 64, 2 residual blocks, and a cross-attention dimensionality of 32. The input image dimensionality constraint was set to 90, 190, and 160.
To mitigate memory costs, we utilized patches of size $64x64x63$ for the Swin-UNETR and $32x32x32$ for the DD-UNET. The networks were trained for 600 epochs with a step learning rate, starting at $5e-3$ and undergoing a reduction by a factor of $0.1$ after 300 epochs. We utilized the Adam optimizer with a weight decay of $1e-4$. In the case of the diffusion model, we implemented an inference of 25 timesteps, following a scaled linear beta schedule with $\beta_{start}=5e-3$ and $\beta_{end}=2e-2$ (\cite{diff}).
\subsection{Fine-tuning methods}
Four distinct fine-tuning methods were employed, as illustrated in Fig \ref{x4}. The first approach, Top-tuning, involved retraining only the top portion of the decoder, encompassing the last 10\% of the model's parameters. The second method, Decoder-Tuning, entailed retraining all the decoder parameters. Full-tuning involved the comprehensive retraining of the entire network. The LoRA approach incorporated linear training from the initial parameters, supplementing them with the pre-training weights of the network, as outlined in \cite{34}. For the purpose of fine-tuning, we introduced an additional layer of an upsampling network to compute the two segmentation labels of interest in the end. In our specific case, these labels correspond to the skeleton sulcal and the PCS region (Fig \ref{x4}). The segmentation loss function, denoted as $loss_{seg}$, utilized a dice loss function instead of the equation \ref{1}, and the total loss function was given by:
\begin{equation}
loss_{total}(x,y)=l_1 loss_{seg}(x,y) +l_2 CE(x,y) 
\label{3}
\end{equation}
where $CE$ is the cross entropy loss function we used for the classification task. After several attempts, $l_1$ and $l_2$ were consistently set to 0.85 and 0.15, respectively, for fine-tuning simulations.
\section{Results}
\subsection{Pre-training simulation}
Fig. \ref{x2} a. visually depicts the outcomes of pre-training simulations, focusing on the self-supervised reconstruction task during pre-training. The simulations revealed that the reconstruction and contrastive learning techniques outperformed adversarial and diffusion methods. The reconstruction task achieved an impressive multi-scale structural similarity score of 55.40\% and a Hausdorff distance of 0.00 mm. For the multi-task classification aspect, adversarial and reconstruction learning yielded classification scores of 0.607 and 0.594, respectively. In terms of memory performance and computational cost, diffusion denoising learning was the most resource-intensive, followed by contrastive and adversarial learning (Fig. \ref{x2} a.). Adversarial learning utilized 51.43\% of the maximum GPU processing power, with an average GPU memory allocation of 97.62\%, requiring 213.4 seconds per epoch for the reconstruction task. The reconstruction task utilized 57.56\% of the maximum GPU power, with an average GPU memory allocation of 95.51\%, and required 179.0 seconds per epoch.  Contrastive learning faced a memory overflow challenge with a batch number of 8, necessitating a reduction to 4, resulting in an increased computation time of 242.2 seconds per epoch. Similarly, diffusion denoising encountered memory overflow issues, leading to adjustments in the region of interest (ROI) size and batch number, resulting in a substantial computational time of 3487 seconds per epoch (Fig. \ref{x2} a. Tab.). The maximum GPU power was reduced to 29.95\%, and the average GPU memory allocation decreased to 55.56\%.
In conclusion, the reconstruction model and adversarial learning emerged as the optimal choices, striking a balance between computational cost and performance in our brain sulci application.
\subsection{Fine-tuning simulations and ablation study}
\begin{figure}
\caption{Results of pre-training and fine-tuning simulations.}
  \label{x2}
      \medskip
      \centerline{ \relax \textbf{(a)} \includegraphics[scale=.35]{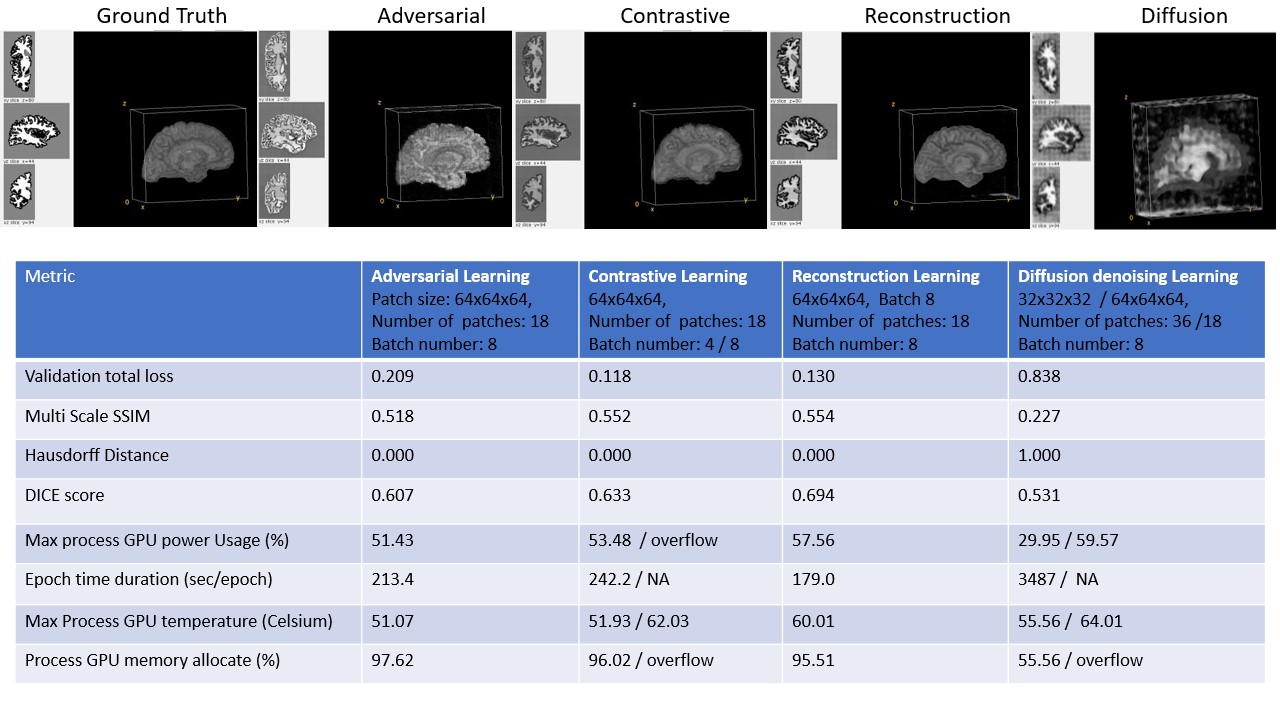}
}
\centerline{
\relax \textbf{(b)}   \includegraphics[scale=.375]{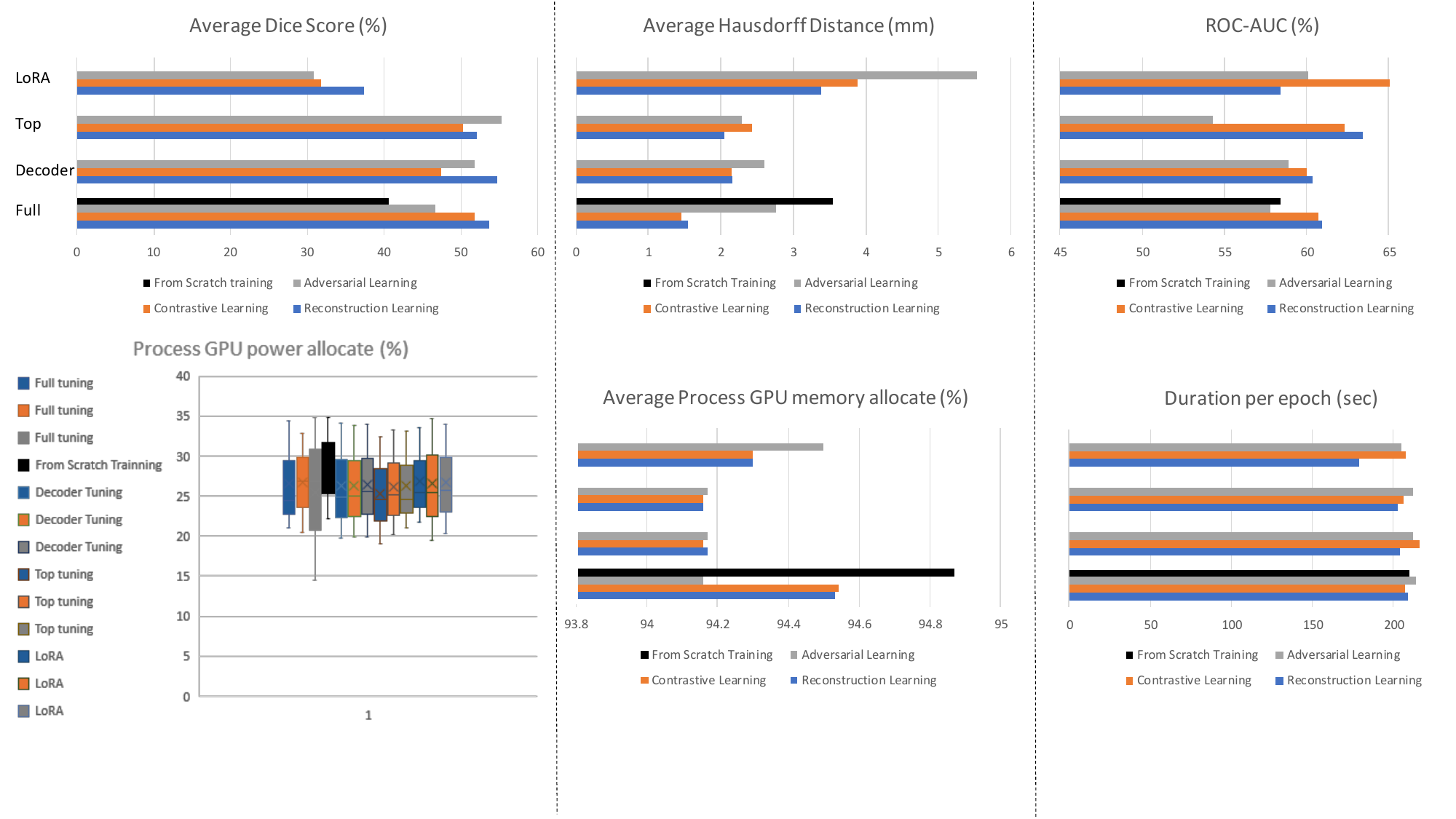}
}
\centerline{
\relax \textbf{(c)}\includegraphics[scale=.375]{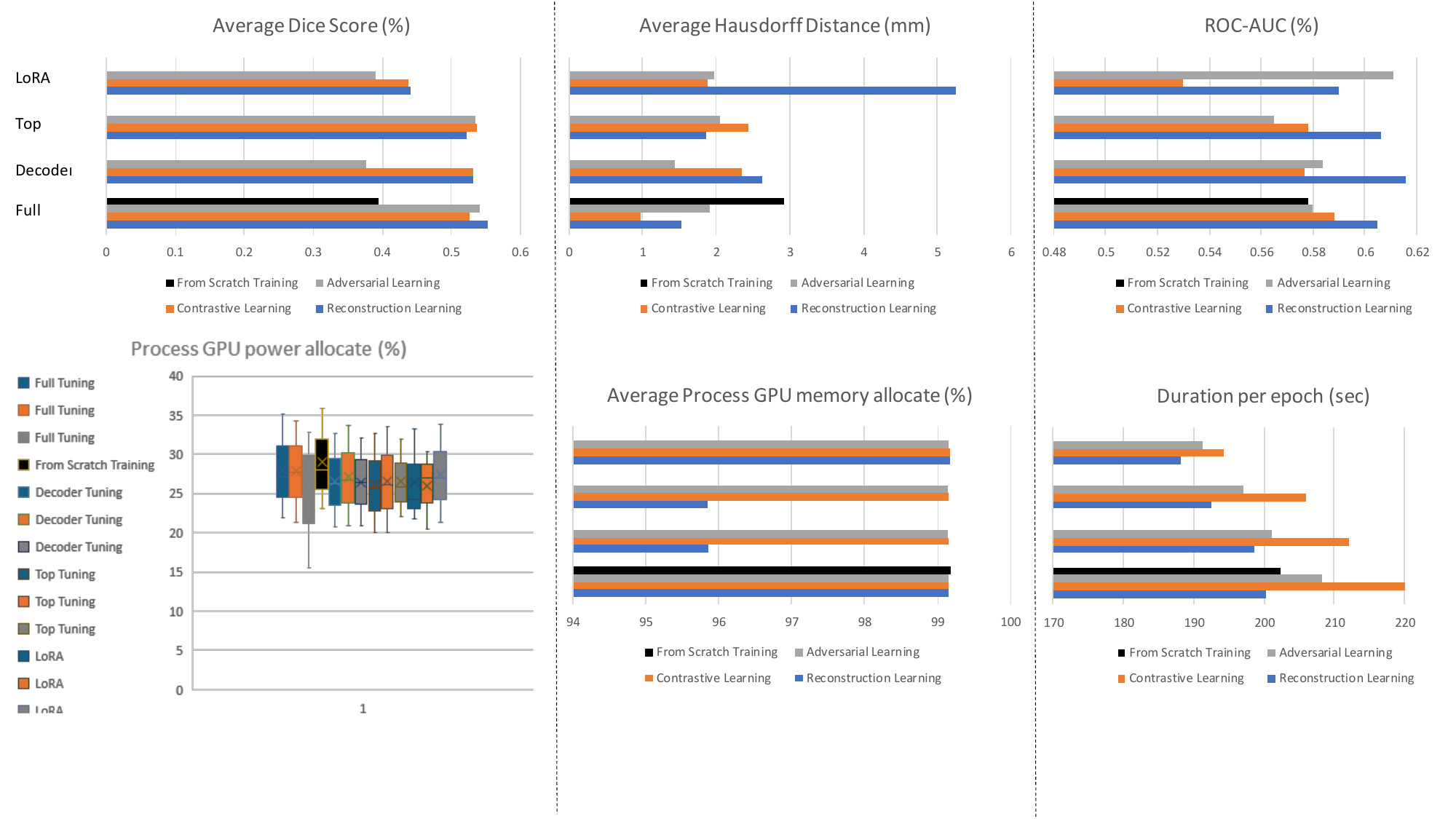}
}
\par\medskip
\textbf{(a-c)} a. Pre-training results of the Swin-Net with a feature size of 24 and DD-UNET (diffusion denoising learning). b. Training the Swin-Net with a feature size of 24. c. Training the Swin-Net with a feature size of 12.
\end{figure}

In our fine-tuning simulations (Fig. \ref{x2} b.) optimal performance was observed with full and top tuning strategies. For the segmentation task, top tuning consistently delivered an average dice score exceeding 50.00\% and an average Hausdorff distance score below 2.50 mm across all pre-training learning simulations. In the classification task, top tuning achieved over 60.00\% for contrastive and reconstruction pre-training, and approximately 54\% for adversarial. Full tuning exhibited similar segmentation performance, surpassing 50.00\% in average dice score and below 1.50 mm in average Hausdorff distance for reconstruction and contrastive learning simulations. For adversarial, it achieved more than 47.00\% in average dice score and less than 2.80 mm in average Hausdorff distance. Decoder tuning closely aligned with full and top tuning, with slightly lower average dice score performance. Low-rank adaptation (LoRA) demonstrated the least favorable results due to the high complexity of the computer vision task, especially in the small region of the PCS. From-scratch training in multi-task learning consistently yielded inferior results compared to top, full, and decoder tuning across various pre-training simulations. Considering computational costs, full tuning emerged as the most resource-intensive, with GPU memory allocation exceeding 94.20\%, process GPU power allocation around 30.00\%, and a duration per epoch surpassing 205 seconds. Conversely, top tuning showcased superior computational efficiency, with GPU memory allocation below 92.20\%, process GPU power allocation around 25.00\%, and a duration per epoch hovering around 200 seconds. LoRA achieved the second most computationally efficient results, closely following the top-tuning approach. From-scratch training incurred the highest computational cost.
\par We utilize an ablation study to compare the performance of two different versions of the Swin-Net model, with a higher parameter count of a 24-feature size and another with fewer parameters of 12-feature size (Fig. \ref{x2} b., c.). Both models exhibited similar trends, with superior performance in full and top tuning and less favorable results in LoRA. The computational cost analysis also echoed the findings of the 24-feature size model, with top tuning be the optimal choice. From-scratch training in the 12-feature size network proved less accurate and computationally expensive, reinforcing the effectiveness of the proposed learning strategy (pre-training and fine-tuning). 
\section{Conclusions}
In this preliminary study, we introduced a strategy that involves both pre-training and fine-tuning to evaluate AI performance in a multi-task IID neuroimaging problem. Our aim was to answer critical questions like: Which pre-training technique—adversarial, contrastive, reconstruction, or diffusion denoising—yields the best results in an IID cohort? How does the performance change with different fine-tuning complexities? Findings highlight the reconstruction model and adversarial learning as the top choices for pre-training, balancing efficiency and performance for our brain sulci application. These insights guide optimal pre-training strategies in multi-task learning. In fine-tuning, full and top tuning outperformed LoRA, emphasizing efficiency in our deep model analysis. Top tuning, particularly, showed peak performance with acceptable computational cost. Our study highlights drawbacks of from-scratch training in brain vision tasks, emphasizing lower accuracy and higher demands. This emphasizes the efficacy of our proposed approach, leveraging pre-training to capture distribution nuances and fine-tuning for precise task execution in intricate multi-task learning within computer vision. Future plans include applying our strategy to diverse cohorts (OOD) and exploring different patch sizes.
\bibliographystyle{splncs04}
\bibliography{library}
\end{document}